\documentclass[twocolumn,superscriptaddress,showpacs,preprintnumbers,prl,
  amsmath,amssymb,final,nofootinbib]{revtex4}

\usepackage{graphicx}
\usepackage{units}
\usepackage{xspace}

\usepackage{pstricks} 

\def\EeV{\ifmmode{\mathrm{Ee\kern -0.1em V}}\else
                   \textrm{Ee\kern -0.1em V}\fi\xspace}%
\def\TeV{\ifmmode{\mathrm{Te\kern -0.1em V}}\else
                   \textrm{Te\kern -0.1em V}\fi\xspace}%
\def\eV{\ifmmode{\mathrm{e\kern -0.1em V}}\else
                   \textrm{e\kern -0.1em V}\fi\xspace}%
\def\gcm{\ifmmode{\mathrm{g/cm}^2}\else
                   {g/cm$^2$}\fi\xspace}%

\newcommand{\maxXlow}{768}
\newcommand{\minXup}{1004}

\newcommand{\etal}{\MakeLowercase{\textit{et al.}\xspace}}

\begin{document}

\title{Measurement of the proton-air cross-section at \boldmath
  $\sqrt{s}=\unit[57]{\TeV}$\\with the Pierre Auger Observatory }


\author{P.~Abreu}
\affiliation{LIP and Instituto Superior T\'{e}cnico, Technical 
University of Lisbon, Portugal}
\author{M.~Aglietta}
\affiliation{Istituto di Fisica dello Spazio Interplanetario 
(INAF), Universit\`{a} di Torino and Sezione INFN, Torino, Italy}
\author{E.J.~Ahn}
\affiliation{Fermilab, Batavia, IL, USA}
\author{I.F.M.~Albuquerque}
\affiliation{Universidade de S\~{a}o Paulo, Instituto de F\'{\i}sica, 
S\~{a}o Paulo, SP, Brazil}
\author{D.~Allard}
\affiliation{Laboratoire AstroParticule et Cosmologie (APC), 
Universit\'{e} Paris 7, CNRS-IN2P3, Paris, France}
\author{I.~Allekotte}
\affiliation{Centro At\'{o}mico Bariloche and Instituto Balseiro 
(CNEA-UNCuyo-CONICET), San Carlos de Bariloche, Argentina}
\author{J.~Allen}
\affiliation{New York University, New York, NY, USA}
\author{P.~Allison}
\affiliation{Ohio State University, Columbus, OH, USA}
\author{A.~Almeda}
\affiliation{Universidad Tecnol\'{o}gica Nacional - Facultad 
Regional Buenos Aires, Buenos Aires, Argentina}
\affiliation{Instituto de Tecnolog\'{\i}as en Detecci\'{o}n y 
Astropart\'{\i}culas (CNEA, CONICET, UNSAM), Buenos Aires, Argentina}
\author{J.~Alvarez Castillo}
\affiliation{Universidad Nacional Autonoma de Mexico, Mexico, 
D.F., Mexico}
% ...
\collaboration{The Pierre Auger Collaboration}
\noaffiliation

\date{\today}

\begin{abstract}
We report a measurement of the proton-air cross-section for particle
production at the center-of-mass energy per nucleon of 57\,\TeV. This
is derived from the distribution of the depths of shower maxima
observed with the Pierre Auger Observatory: systematic uncertainties
are studied in detail.  Analysing the tail of the distribution of the
shower maxima, a proton-air cross-section of
$\left[505\;\pm22(\text{stat})\;^{+28}_{-36}(\text{sys})\right]\,$mb
is found.
\end{abstract}


\maketitle

\section{Introduction}

We present the first analysis of the proton-air cross-section based on
measurements made at the Pierre Auger
Observatory~\cite{Abraham:2004dt}.  For this purpose we analyse the
shape of the distribution of the largest values of the depth of shower
maximum, $X_{\rm max}$, the position at which an air shower deposits
the maximum energy per unit of mass of atmosphere traversed.  The
\emph{tail} of the $X_{\rm max}$-distribution is sensitive to the
proton-air cross-section, a fact first exploited in the pioneering
work of the Fly's Eye Collaboration~\cite{Ellsworth:1982kv}.  To
obtain accurate measurements of $X_{\rm max}$, timing data from the
fluorescence telescopes is combined with that from the surface
detector array for a precise hybrid reconstruction of the geometry of
events~\cite{Sommers:1995dm}.

We place particular emphasis on studying systematic uncertainties in
the cross-section analysis.  The unknown mass composition of
cosmic-rays~\cite{PedroICRC} is identified to be \textit{the} major
source of systematic uncertainty and accordingly the analysis has been
optimised to minimise the impact of particles other than protons in
the primary beam. This begins with restricting the analysis to the
energy interval $10^{18}$ to $10^{18.5}\,$\eV, where the shape of the
$X_{\rm max}$-distribution is compatible with there being a
substantial fraction of protons; also there are a large number of
events recorded in this energy range. The corresponding average
center-of-mass energy of a proton interacting with a
nucleon is \unit[57]{\TeV}, significantly above the
reach of the Large Hadron Collider.

\section{Analysis Approach}

The proton-air cross-section is derived in a two-step process.
Firstly, we measure an air-shower observable with high sensitivity to
the cross-section.  Secondly, we convert this measurement to a value
of the proton-air cross-section for particle production (c.f.\ \cite{Engel:1998pw}).
This is the cross-section that accounts for all interactions which produce particles and thus
contribute to the air-shower development; it implicitly also
includes diffractive interactions. As the primary observable we
define $\Lambda_\eta$ via the exponential shape of the tail of the
$X_{\rm max}$-distribution, ${\rm d}N/{\rm d}X_{\rm max}\propto
\exp(-X_{\rm max}/\Lambda_\eta)$, where $\eta$ denotes the fraction of
most deeply penetrating air showers used.  Considering only these
events enhances the contribution of protons in the sample, since the
depth at which proton-induced showers maximise is deeper in the
atmosphere than for showers from heavier nuclei. Thus, $\eta$ is a key
parameter: a small value enhances the proton fraction, but reduces the
number of events available for the analysis. We have chosen $\eta=0.2$
so that, for helium-fractions up to 25\,\%, biases introduced by the
possible presence of helium and heavier nuclei do not exceed the level
of the statistical uncertainty. This was chosen after a Monte Carlo
study that probed the sensitivity of the analysis to the mass
composition depending on the choice of different values of~$\eta$.

\section{The Measurement of \boldmath$\Lambda_\eta$}

We use events collected between 1 Dec 2004 and 20 Sept 2010.  The
atmospheric and event-quality cuts applied are identical to those used
for the analysis of $\langle X_{\rm max}\rangle$ and RMS($X_{\rm
  max}$)~\cite{Abraham:2010yv} yielding $11\kern0.1em628$ high-quality
events.
The $X_{\rm max}$-distribution of these data is affected by the
known geometrical acceptance of the fluorescence telescopes as well as
by limitations related to atmospheric light transmission.  We use the
strategy developed for the measurement of $\langle X_{\rm
  max}\rangle$ and RMS($X_{\rm max}$) to extract a sample that has an
unbiased $X_{\rm max}$-distribution: a fiducial volume selection, which
requires event geometries that allow, for each individual shower, the
complete observation of a defined slant depth range.

Firstly, we derive the range of values of $X_{\rm max}$ that
corresponds to the fraction $\eta=0.2$ of the most deeply penetrating
showers. For this we need an unbiased distribution of $X_{\rm max}$
over the entire depth range of observed values of $X_{\rm max}$. To
achieve this we perform a fiducial event selection of the slant depth
range containing 99.8\,\% of the observed $X_{\rm max}$-distribution,
which corresponds to the range from $550$ to
$\unit[1004]{g/cm^{2}}$. This reduces the data sample to $1635$ events
providing an unbiased $X_{\rm max}$-distribution that is used to find
the range of values of $X_{\rm max}$ corresponding to $\eta=0.2$,
identified to extend from \maxXlow\xspace to \minXup\,\gcm.

\begin{figure}[t]
  \includegraphics[width=\linewidth]{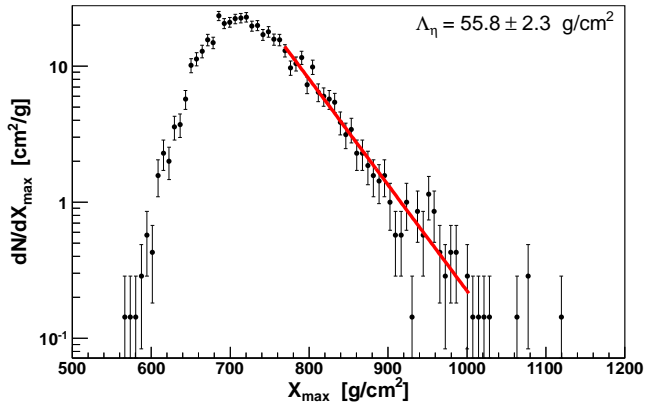}
  \caption{\label{fig:fit} Unbinned likelihood fit to obtain
    $\Lambda_\eta$ (thick line). The $X_{\rm max}$-distribution is unbiased by the
    fiducial geometry selection applied in the range of the fit. }
\end{figure}

Secondly, we select those events from the original $11\kern0.1em628$
that have geometries allowing the complete observation of values of
$X_{\rm max}$ from \maxXlow\xspace to \minXup\,\gcm, the tail of the
unbiased distribution. This fiducial cut maximises the statistics of
an unbiased $X_{\rm max}$-distribution in the range of interest.  In
total $3082$ events pass the fiducial volume cuts, of which $783$
events have their $X_{\rm max}$ in the selected range and thus
contribute directly to the measurement of $\Lambda_\eta$.  In
Fig.~\ref{fig:fit} we show the $3082$ selected events and the result
of an unbinned maximum likelihood fit of an exponential function over
the range $768$ to $\unit[1004]{g/cm^2}$.  Values of $\Lambda_\eta$
have been re-calculated for sub-samples of the full dataset selected
according to zenith-angle, shower-to-telescope distance and energy: the
different values obtained for $\Lambda_\eta$ are consistent with
statistical fluctuations. The re-analyses of the data for changes of
fiducial event selection, modified values of $\eta$ and for different
ranges of atmospheric depths yield changes of $\Lambda_\eta$ that are
distributed around zero with a root-mean-square of
$\unit[1.6]{\gcm}$. We use this root-mean-square as an estimate of the
systematic uncertainties associated to the measurement. This yields
\begin{equation}
  \Lambda_\eta = \left[55.8\; \pm 2.3(\text{stat})\; \pm 1.6(\text{sys})\right]\,\gcm,
\end{equation}
with the average energy of these events being
$10^{18.24\;\pm0.005(\text{stat})}\,$\eV. The differential energy distribution for these events follows a
power-law with index $-1.9$. The average energy corresponds to a
center-of-mass energy of $\sqrt{s}=\left[57\;\pm0.3(\text{stat})\right]\,$\TeV{} in proton-proton
collisions.

\section{Determination of the Cross-Section}

The determination of the proton-air cross-section for particle
production requires the use of air-shower simulations, which
inherently introduces some dependence on model assumptions.  We
emulate the measurement of $\Lambda_\eta$ with Monte Carlo simulations
to derive predictions of the slope, $\Lambda_\eta^{\rm MC}$.  It is
known from previous work that the values of $\Lambda_\eta^{\rm MC}$
are directly linked to the hadronic cross-sections used in the
simulations~\cite{Ellsworth:1982kv}.  Accordingly we can explore the
effect of changing cross-sections empirically by multiplying all
hadronic cross-sections input to the simulations by an
energy-dependent factor~\cite{Ulrich:2009zq}
\begin{equation}
  f(E,\,f_{19}) = 1 + (f_{19} - 1)\; \dfrac{\lg\left(E/\unit[10^{15}]{\eV}\right)}{\lg\left(\unit[10^{19}]{\eV}/\unit[10^{15}]{\eV}\right)},
  \label{eq:factor}
\end{equation}
where $E$ denotes the shower energy and $f_{19}$ is the factor by
which the cross-section is rescaled at $\unit[10^{19}]{\eV}$. This
factor is unity below $\unit[10^{15}]{\eV}$ reflecting the fact that
measurements of the cross-section at the Tevatron were used to tune
the interaction models.  This technique of modifying the original
predictions of the cross-sections during the simulation process
assures a smooth transition from accelerator data up to the energies
of our analysis.

For each hadronic interaction model, the value of $f_{19}$ is obtained
that reproduces the measured value of $\Lambda_\eta$. The modified
cross-section is then deduced by multiplying the original
cross-section used in the model by the factor $f(E,\,f_{19})$ of
Eq.~(\ref{eq:factor}) using $E=10^{18.24}\,$\eV.  For the conversion of
$\Lambda_\eta$ into cross-section, we have used the four high-energy
hadronic interaction models commonly adopted for air shower
simulations: QGSJet01~\cite{Kalmykov:1993qe},
QGSJetII.3~\cite{Ostapchenko:2005nj}, SIBYLL~2.1~\cite{Ahn:2009wx} and
EPOS1.99~\cite{Werner:2005jf}. While in general no model gives a
completely accurate representation of cosmic-ray data in all respects,
these have been found to give reasonably good descriptions of many of
the main features. 
{It has been shown~\cite{Parsons:2011ad} that the differences between the
models used in the analysis are typically bigger than the variations
obtained within one model by parameter variation. Therefore we use the
model differences for estimating the systematic model dependence.
}

The proton-air cross-sections for particle production derived for
QGSJet01, QGSJetII, SIBYLL and EPOS are $523.7$,
$502.9$, $496.7$ and $497.7$\,mb respectively,
with the statistical uncertainty for each of these values being $22\,$mb.
The difference of these cross-sections from the original model
predictions are $<5$\,\%, with the exception of the result obtained
with the SIBYLL model, which is $12\,\%$ smaller than the original
SIBYLL prediction. 
We use the maximum deviations derived from using
the four models, relative to the average result of $505\,$mb, to
estimate a systematic uncertainty of $(-8,\,+19)\,$mb related to the
difficulties of modelling high energy interactions.
This procedure relies on the coverage of the underlying theoretical
uncertainties by the available models.  For example diffraction,
fragmentation, inelastic intermediate states, nuclear effects, QCD
saturation, etc.\ are all described at different levels 
using different phenomenological, but self-consistent, approaches 
in these models.  It is thus possible
that the true range of the uncertainties for air-shower analyses is
larger, but this cannot be estimated with these models.
Furthermore,
certain features of hadronic particle production,
such as the multiplicity, elasticity and pion-charge ratio, have an
especially important impact on air shower
development~\cite{Matthews:2005sd,Ulrich:2010rg}; of these we found
that only the elasticity can have a relevant impact on $\Lambda_\eta$.
{The previously identified systematic uncertainty of $(-8,\,+19)\,$mb
induced by the modelling of hadronic interactions,
corresponds to the impact of modifying the elasticity
within $\pm(10-25)\,\%$ in the models.}

The selection of events with large values of $X_{\rm max}$ also
enhances the fraction of primary cosmic-ray interactions with smaller
multiplicities and larger elasticities, which is for example
characteristic for diffractive interactions. The value of
$\Lambda_\eta$ is thus more sensitive to the cross-section of those
interactions. The identified model-dependence for the determination of
$\sigma^{\rm prod}_{p\kern 0.1em\text{-air}}$ is also caused by the
compensation of this effect.

Also the choice of a logarithmic energy dependence for the
rescaling-factor in Eq.~(\ref{eq:factor}) may affect the resulting
cross-sections. However, since the required rescaling-factors are
small, this can only be a marginal effect.

The systematic uncertainty of 22\,\%~\cite{EThis} in the absolute
value of the energy scale leads to systematic uncertainties of 7\,mb
in the cross-section and $6\,$\TeV in the center-of-mass energy.
Furthermore, the procedure to obtain $\sigma^{\rm prod}_{p\kern
  0.1em\text{-air}}$ from the measured $\Lambda_\eta$ depends on
additional parameters. By varying the energy distribution, energy and
$X_{\rm max}$ resolution in the simulations, we find that related
systematic changes of the value of $\sigma^{\rm prod}_{p\kern
  0.1em\text{-air}}$ are distributed with a root-mean-square of
$\unit[7]{mb}$ around zero. We use the root-mean-square as estimate of
the systematic uncertainties related to the conversion of
$\Lambda_\eta$ to $\sigma^{\rm prod}_{p\kern 0.1em\text{-air}}$.

\begin{table}[b]
  \caption{\label{tab:syst}Summary of the systematic uncertainties.}
  \vspace*{.2cm}
  \centering
  \begin{tabular}{lr}
    Description & Impact on $\sigma^{\rm prod}_{p\kern 0.1em\text{-air}}$\\
    \hline \hline
    $\Lambda_\eta$ systematics & $\pm15\,$mb \\
    Hadronic interaction models & $_{-8}^{+19}$\,mb\\
    Energy scale & $\pm7\,$mb \\
    Conversion of $\Lambda_\eta$ to $\sigma^{\rm prod}_{p\kern
  0.1em\text{-air}}$ & $\pm7\,$mb \\
    \hline
    Photons, $<$0.5\,\% & $<+10$\,mb \\
    Helium, 10\,\% & $-12\,$mb \\
    Helium, 25\,\% & $-30\,$mb \\
    Helium, 50\,\% & $-80\,$mb \\
    \hline
    Total (25\,\% helium) & $-36\,$mb, $+28\,$mb\\
  \end{tabular}
\end{table}

The presence of photons in the primary beam would bias the
measurement.  The average $X_{\rm max}$ of showers produced
by photons at the energies of interest is about $\unit[50]{g/cm^2}$
deeper in the atmosphere than that of protons.  However, observational
limits on the fraction of photons are
$<0.5$\,\%~\cite{Glushkov:2009tn,PhThis}. With simulations we find
that the possible under-estimation of the cross-section if photons
were present in the data sample {at this level} is less than 10\,mb.

\begin{figure}[t]
  \includegraphics[width=\linewidth]{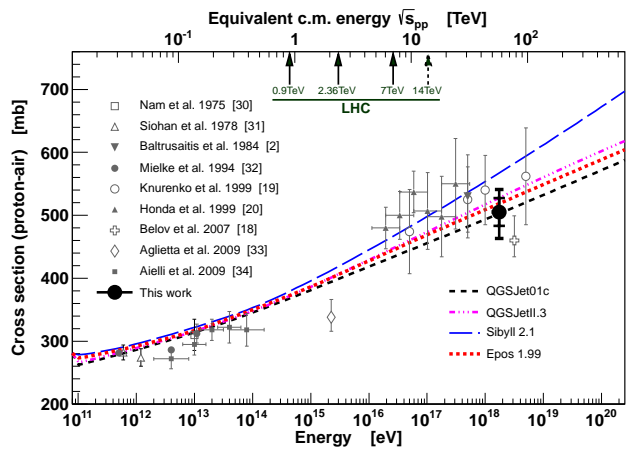}
  \caption{\label{fig:result}Resulting $\sigma^{\rm prod}_{p\kern
      0.1em\text{-air}}$ compared to other
    measurements~(see~\cite{Nam:1975xk,Siohan:1978zk,Mielke94,%
      Honda:1993kv,Knurenko:1999cr,Aglietta:2009zz,Belov:2006mb,Aielli:2009ca}
    for references) and model predictions. The inner error bars are
    statistical, while the outer include systematic uncertainties for
    a helium fraction of 25\,\% and 10\,mb for the systematic
    uncertainty attributed to the fraction of photons.}
\end{figure}

With the present limitations of observations, we cannot
distinguish air showers produced by helium nuclei from those created by
protons.  From simulations we find that $\sigma^{\rm prod}_{p\kern 0.1em\text{-air}}$ is
over-estimated depending on the percentages of helium in the data
sample. Lack of knowledge of the helium fraction is the dominant
source of systematic uncertainty.

We also find that
the nuclei of the CNO-group introduce no bias for fractions up to
$\sim50$\,\%, and accordingly we assign no uncertainty in the
cross-section due to these or heavier nuclei.

In Table~\ref{tab:syst} we list the sources of systematic
uncertainties.  As the helium fraction is not known we show the impact
of 10, 25 and 50\,\% of helium respectively.
In what follows we include a systematic uncertainty related to a helium
fraction of 25\,\%. In the extreme case, were the
cosmic-ray composition to be 100\,\% helium, the analysis would
over-estimate the proton-air cross-section by 300 to 500\,mb. Given the
constraints from accelerator data at lower energies and typical model
assumptions, this extreme scenario is not realistic.

We summarise our results by averaging the four values of the
cross-section obtained with the hadronic interaction models to give
\begin{equation*}
  \sigma^{\rm prod}_{p\kern 0.1em\text{-air}} = \left[505\;\pm 22(\text{stat})\;
  _{-36}^{+28}(\text{sys})\right] \,\rm{mb}
\end{equation*}
at a center-of-mass energy of
$\left[\right.57\;\pm0.3(\text{stat})\;\pm6(\text{sys})\left.\right]$\,\TeV.
In Fig.~\ref{fig:result} we compare this result with model predictions
and other measurements.  The measurements at the highest energies are:
HiRes~\cite{Belov:2006mb} and Fly's eye~\cite{Ellsworth:1982kv} that
are both based on $X_{\rm max}$, Yakutsk Array~\cite{Knurenko:1999cr}
using Cherenkov observations and Akeno~\cite{Honda:1993kv} measuring
electron and muon numbers at ground level. All these analyses assume a
pure proton composition. In the context of a
possible mixed-mass cosmic-ray composition, this can lead to large
systematic effects. Also all these analyses are based on a single
interaction model for describing air showers: Only HiRes 
uses a second model for systematic checks.

It is one of the prime aims of our analysis to have the smallest
possible sensitivity to a non-proton component, and to perform a
detailed systematic analysis on the uncertainties related to the mass
composition. We also use, for the first time, all hadronic interaction
models currently available for the estimation of model-related
systematic effects. Futhermore, by using Eq.~(\ref{eq:factor}) we
derive a cross-section corresponding to a smooth interpolation from
the Tevatron measurement to our analysis, with no inconsistencies as
in earlier approaches.


\begin{figure}[b]
  \includegraphics[width=\linewidth]{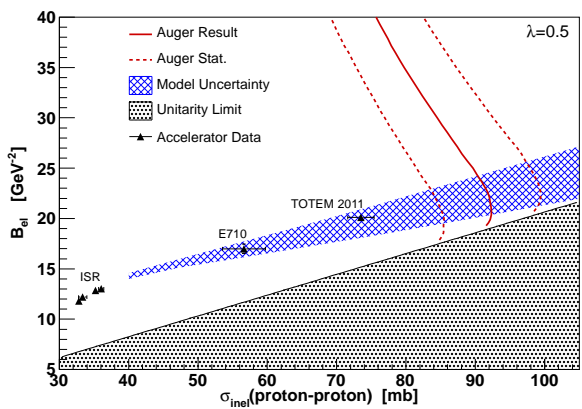}
  \caption{\label{fig:correlation}Correlation of elastic slope
    parameter, $B_{\rm el}$, and the inelastic proton-proton
    cross-section in the Glauber framework. The solid line indicates
    the parameter combinations yielding the observed proton-air
    production cross-section, and the dotted lines are the statistical
    uncertainties. The hatched area corresponds to the predictions by
    SIBYLL, QGSJet, QGSJetII and EPOS. See also
    Ref.~\cite{Engel:1998pw}.}
\end{figure}

\section{Comparison With Accelerator Data}

{
For the purpose of making comparisons with accelerator data we 
 calculate the inelastic and total proton-proton
  cross-sections using the Glauber model.  We use standard Glauber
  formalism~\cite{Glauber:1955qq},
%
%
 extended by a two-channel
  implementation of inelastic intermediate states~\cite{Kalmykov:1993qe} to account for
  diffraction dissociation~\cite{GoodWalker1960}. The first channel
  corresponds to $p\rightarrow p$ scattering and has an amplitude of
  $\Gamma_{pp}$, while the amplitude for the other channel is
  $\Gamma_{pp^*}=\lambda\,\Gamma_{pp}$ and corresponds to the
  excitation of a short lived intermediate state. The parameter
  $\lambda$ is related to the ratio of single-diffractive
  cross-section and elastic cross-section. We use a value of
  $\lambda=0.5\pm0.15$ that is determined from measurements of the
  single-diffractive cross-section, as well as from proton-carbon cross-section
  data at lower energies.}

This Glauber calculation is
model-dependent since neither the parameters nor the physical
processes involved are known accurately at cosmic-ray energies. 
{In particular, this applies to} the elastic slope parameter,
$B_{\rm el}$, {(defined
by ${\rm d}\sigma_{\rm el}/{\rm d}t\propto\exp(-|t|B_{\rm el})$ 
for very small $t$)}, 
the correlation of $B_{\rm el}$ to the cross-section, 
{and the cross-section
for diffractive dissociation.}
For the example of $\sigma_{pp}^{\rm inel}$,
the correlation of $B_{\rm el}$ with the cross-section is
shown in Fig.~\ref{fig:correlation} {for $\lambda=0.5$}.  We have used the same four hadronic
interaction models to determine the uncertainty band of the $B_{\rm
  el}-\sigma_{pp}^{\rm inel}$ correlation.  Recent cross-section
models such as~\cite{Block:2006hy} fall within this band.  We find
that in the Glauber framework the \textit{inelastic} cross-section is
less dependent on model assumptions than the \textit{total}
cross-section.  The result for the inelastic proton-proton
cross-section is {
\begin{equation*}
  \sigma_{pp}^{\rm inel} = \left[92 \;\pm 7(\text{stat}) \;
  _{-11}^{+9}(\text{sys}) \;\pm 7(\text{Glauber})\right]\,\rm{mb},
\end{equation*}
}
and the total proton-proton cross-section is {
\begin{equation*}
  \sigma_{pp}^{\rm tot} = \left[133 \;\pm 13(\text{stat})\;
  _{-20}^{+17}(\text{sys}) \;\pm 16(\text{Glauber})\right] \,\rm{mb}.
\end{equation*}
The systematic uncertainties for the inelastic and total
cross-sections include contributions from the elastic slope parameter,
from $\lambda$, from the description of the nuclear density profile,
and from cross-checking these effects using
QGSJetII~\cite{Ostapchenko:2005nj,Ostapchenko:2010gt}.  For the
inelastic case, these three independent contributions are 1, 3, 5, and
4\,mb respectively.  For the total cross-section, they are 13, 6, 5,
and 4\,mb.}  {We emphasize that the total theoretical uncertainty
  of converting the proton-air to a proton-proton cross-section may be
  larger than estimated here within the Glauber model. There are other
  extensions of the Glauber model to account for inelastic
  screening~\cite{Kalmykov:1993qe,Guzey:2005tk} or nucleon-nucleon
  correlations \cite{Baym:1995cz}, and alternative approaches that
  include, for example, parton saturation or other effects
  \cite{Werner:2005jf,Ostapchenko:2010gt,Frankfurt93a,Harrington:2002gc}.
}

\begin{figure}[tb]
  \includegraphics[width=\linewidth]{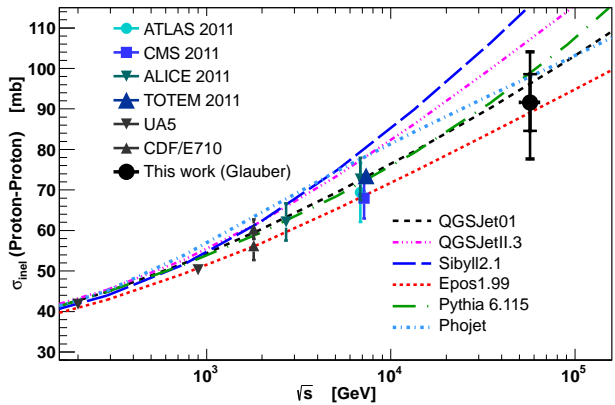}
  \caption{\label{fig:pp} Comparison of derived $\sigma_{pp}^{\rm
      inel}$ to model predictions and {accelerator data~\cite{lhc}}.
    Here we also show the cross-sections of two typical high-energy
    models, Pythia6~\cite{Schuler:1993wr} and Phojet\cite{Engel:1994vs}.  The inner
    error bars are statistical, while the outer include systematic
    uncertainties.}
\end{figure}

In Fig.~\ref{fig:pp} we compare our inelastic cross-section result
to accelerator data and to the cross-sections used in the hadronic
interaction models.

\section{Summary}

We have presented the first measurement of the cross-section for the
production of particles in proton-air collisions from data collected
at the Pierre Auger Observatory.  We have studied in detail the effects of
assumptions on the primary cosmic-ray mass composition, hadronic
interaction models, simulation settings and the fiducial volume limits
of the telescopes on the final result.  By analysing only
the most deeply penetrating events we selected a data sample enriched in
protons. The results are presented assuming a maximum contamination of
25\,\% of helium in the light cosmic-ray mass component.
The lack of knowledge of the helium component is
the largest source of systematic uncertainty.  However, for helium
fractions up to 25\,\% the induced bias remains below $6\,$\%.

To derive a value of $\sigma^{\rm prod}_{p\kern 0.1em\text{-air}}$
from the measured $\Lambda_\eta$ we assume a smooth extrapolation of
hadronic cross-sections from accelerator measurements to the energy of
the analysis. This is achieved by modifying the model-predictions of
hadronic cross-sections above energies of $10^{15}\,\eV$ during the
air-shower simulation process in a self-consistent approach.

We convert the proton-air production cross-section into the total, and
the inelastic, proton-proton cross-section using a Glauber calculation
{that includes intermediate inelastic screening corrections}. In
this calculation we use the correlation between the elastic slope
parameter and the proton-proton cross-sections taken from the
interaction models as a constraint.  We find that the inelastic
proton-proton cross-section depends less on the elastic slope
parameter than does the total proton-proton cross-section, and thus
the systematic uncertainty of the Glauber calculation for the
inelastic result is smaller. 
{The data agree with an extrapolation from LHC~\cite{lhc} 
energies to 57\,TeV for a limited set of models.}

\noindent\textit{Acknowledgments}.
The successful installation, commissioning, and operation of the Pierre Auger Observatory
would not have been possible without the strong commitment and effort
from the technical and administrative staff in Malarg\"ue.

We are very grateful to the following agencies and organizations for financial support:
Comisi\'on Nacional de Energ\'ia At\'omica,
Fundaci\'on Antorchas,
Gobierno De La Provincia de Mendoza,
Municipalidad de Malarg\"ue,
NDM Holdings and Valle Las Le\~nas, in gratitude for their continuing
cooperation over land access, Argentina;
the Australian Research Council;
Conselho Nacional de Desenvolvimento Cient\'ifico e Tecnol\'ogico (CNPq),
Financiadora de Estudos e Projetos (FINEP),
Funda\c{c}\~ao de Amparo \`a Pesquisa do Estado de Rio de Janeiro (FAPERJ),
Funda\c{c}\~ao de Amparo \`a Pesquisa do Estado de S\~ao Paulo (FAPESP),
Minist\'erio de Ci\^{e}ncia e Tecnologia (MCT), Brazil;
AVCR AV0Z10100502 and AV0Z10100522,
GAAV KJB100100904,
MSMT-CR LA08016, LC527, 1M06002, MEB111003, and MSM0021620859, Czech Republic;
Centre de Calcul IN2P3/CNRS,
Centre National de la Recherche Scientifique (CNRS),
Conseil R\'egional Ile-de-France,
D\'epartement  Physique Nucl\'eaire et Corpusculaire (PNC-IN2P3/CNRS),
D\'epartement Sciences de l'Univers (SDU-INSU/CNRS), France;
Bundesministerium f\"ur Bildung und Forschung (BMBF),
Deutsche Forschungsgemeinschaft (DFG),
Finanzministerium Baden-W\"urttemberg,
Helmholtz-Gemeinschaft Deutscher Forschungszentren (HGF),
Ministerium f\"ur Wissenschaft und Forschung, Nordrhein-Westfalen,
Ministerium f\"ur Wissenschaft, Forschung und Kunst, Baden-W\"urttemberg, Germany;
Istituto Nazionale di Fisica Nucleare (INFN),
Ministero dell'Istruzione, dell'Universit\`a e della Ricerca (MIUR), Italy;
Consejo Nacional de Ciencia y Tecnolog\'ia (CONACYT), Mexico;
Ministerie van Onderwijs, Cultuur en Wetenschap,
Nederlandse Organisatie voor Wetenschappelijk Onderzoek (NWO),
Stichting voor Fundamenteel Onderzoek der Materie (FOM), Netherlands;
Ministry of Science and Higher Education,
Grant Nos. N N202 200239 and N N202 207238, Poland;
Funda\c{c}\~ao para a Ci\^{e}ncia e a Tecnologia, Portugal;
Ministry for Higher Education, Science, and Technology,
Slovenian Research Agency, Slovenia;
Comunidad de Madrid,
Consejer\'ia de Educaci\'on de la Comunidad de Castilla La Mancha,
FEDER funds,
Ministerio de Ciencia e Innovaci\'on and Consolider-Ingenio 2010 (CPAN),
Xunta de Galicia, Spain;
Science and Technology Facilities Council, United Kingdom;
Department of Energy, Contract Nos. DE-AC02-07CH11359, DE-FR02-04ER41300,
National Science Foundation, Grant No. 0450696,
The Grainger Foundation USA;
ALFA-EC / HELEN,
European Union 6th Framework Program,
Grant No. MEIF-CT-2005-025057,
European Union 7th Framework Program, Grant No. PIEF-GA-2008-220240,
and UNESCO.


\end{document}